\documentclass[preprint,12pt]{revtex4-1}

\usepackage{graphicx}
\usepackage{dcolumn}
\usepackage{epsfig} 
\usepackage{bm}
\usepackage{epstopdf}
\usepackage{amssymb}
\usepackage{color}

\def\be{\begin{equation}}
\def\bea{\begin{eqnarray}}
\def\ee{\end{equation}}
\def\eea{\end{eqnarray}}

\def\pt{\partial}

\def\dt{\delta}

\def\ffi{\varphi}

\def\La{\Lambda}

\def\dd{\mbox{d}}

\def\const{\mbox{const}}

\newcommand{\planss}  {{Planatary Space Science }}  
\newcommand{\ssr}{    {Space Sci. Rev. }}
\newcommand{\grl}{    {Geophys Res. Lett. }}
\newcommand{\jgr}{    {J. Geophys. Res. }}

\begin{document}


\title{Kinetic equation for nonlinear wave-particle interaction: solution properties and asymptotic dynamics.}




\author{A. V. Artemyev}
 \altaffiliation[Also at ]{Space Research Institute, RAS, Moscow, Russia}
\email{aartemyev@igpp.ucla.edu}
\affiliation{
Institute of Geophysics and Planetary Physics, UCLA, Los Angeles, California, USA.
}%

\author{A. I. Neishtadt}
 \altaffiliation[Also at ]{Space Research Institute, RAS, Moscow, Russia}
\affiliation{
Department of Mathematical Sciences, Loughborough University, Loughborough LE11 3TU, UK.
}%

\author{A. A. Vasiliev}
\affiliation{Space Research Institute, RAS, Moscow, Russia.}
\email{Corresponding author. E-mail: valex@iki.rssi.ru}

\date{\today}

\begin{abstract}
We consider a kinetic equation describing evolution of a particle distribution function in a system with nonlinear wave-particle interactions (trappings into a resonance and nonlinear scatterings). We study properties of its solutions and show that the only stationary solution is a constant, and that all solutions with smooth initial conditions tend to constant as time grows. The resulting flattening of the distribution function in the domain of nonlinear interactions is similar to one described by the quasi-linear plasma theory, but the distribution evolves much faster. The results are confirmed numerically for a model problem.
\end{abstract}




\maketitle

\section{Introduction}
In sufficiently dense plasma (a gas of ionized particles), the energy exchange between particles and the energy transformation (e.g., from kinetic to thermal energy) are controlled by particle collisions, which lead to their momentum exchange \cite[e.g.,][]{Spitzer62}. Rarefied space plasma systems, however, are collisionless, and hence other kinetic processes control the energy exchange and dissipation within them. Electromagnetic waves generated by one population of particles can travel in space and interact with another one, possibly quite distant, particle population. This interaction effectively connects particles which never physically collide with each other, and results in collisionless momentum exchanges. Therefore, understanding of the thermalization process in collisionless plasma requires a detailed picture of the wave-particle interaction \cite[e.g.,][]{Galeev&Sagdeev79,Gary:book05}.

Wave-particle resonant interaction plays a crucial role in the structure and dynamics of various space and laboratory plasma systems. In the near-Earth space environment, this interaction determines structure and dynamics of the collisionless bow shock \cite[e.g.,][and references therein]{Krasnoselskikh13}, controls solar wind transport across the Earth magnetosphere boundary -- magnetopause \cite[e.g.,][and references therein]{Wing14}, participates in magnetic energy release in the magnetotail reconnection \cite[e.g.,][and references therein]{Fujimoto11}, contributes to field-aligned current dissipation \cite[e.g.,][and references therein]{Lysak90}, determines the radiation belt formation and dynamics \cite[e.g.,][and references therein]{Thorne10:GRL}, controls charged particle precipitation to aurora \cite[e.g.,][and references therein]{Ni16:ssr} and their acceleration in the aurora region \cite[e.g.,][and references therein]{Watt&Rankin12}.

The basic concept describing wave-particle interaction is the quasi-linear theory proposed in early 60s \cite{Drummond&Pines62,Vedenov62}, based on the assumption of charged particle stochastic scattering by plasma turbulence with sufficiently broad spectrum. However, as high-resolution wave measurements become available \cite[e.g.,][]{Matsumoto94, Panov06, Bale&Mozer07, Cattell08,Cully08, Khotyaintsev10, Chaston13, Wilder16}, the applicability of quasi-linear theory becomes questionable. Intense coherent waves at the bow shock and in the Earth's magnetosphere can lead to effective momentum exchanges between particle populations and support anomalous plasma transport and energy dissipation. However, their resonant interaction with charged particles is nonlinear and cannot be described by the quasi-linear theory. This problem stimulates the development of new models and theories of wave-particle resonant interactions and their contribution to large-scale plasma system structure and dynamics \cite[e.g.,][]{Omura15, Osmane16, Hsieh&Omura17, Shklyar17, Demekhov17, Artemyev18:cnsns}.

One of the perspective models of nonlinear wave-particle interaction consists in generalization of the kinetic equation (generalized Fokker-Planck equation) in order to include effects of particle nonlinear scattering (non-diffusive drift in phase space \cite[see, e.g.,][]{Shklyar09:review, Artemyev14:pop, Albert13:AGU}) and nonlocal transport (large jumps in the phase space due to phase trapping effect \cite[see, e.g.,][]{Omura07, Demekhov09, Artemyev16:ssr}). The combination of the Hamiltonian theory of perturbations in the resonant systems \cite{bookAKN06} and the probabilistic approach for describing resonant systems results in such a kinetic equation (generalized Fokker-Planck equation) incorporating non-diffusive effects of nonlinear wave-particle interaction. This equation successfully describes the evolution of charged particle distribution in the different space plasma systems and explains many important effects observed by spacecraft in the near-Earth plasma environment \cite[see, e.g., discussion in][]{Artemyev18:jpp, Zhang18:jgr:intensewaves, Mourenas18:jgr}. In this paper we investigate general properties of this kinetic equations and its solutions. In Section 2 we demonstrate existence of a unique stationary solution of the kinetic equation, which is a constant solution. In Section 3 we consider a simplified version of the equation and construct its general solution; then we show that for any smooth initial distribution it tends to a constant solution as time tends to infinity. In section 4 we use numerical simulations of the complete equation to test our analytical results derived in Sections 2 and 3. And finally in Section 5 we discuss the paper conclusions.

\section{The kinetic equation and its properties}

In \cite{Artemyev16:pop:letter,Artemyev17:pre} (see also \cite{Artemyev17:arXiv}) we have introduced the kinetic equation describing evolution of the distribution function of charged particles in a system with captures (trapping) into a resonance and scatterings on the resonance. The distribution function $f$ depends on the action variable $I$ (a function of the particle energy, see \cite{Artemyev17:pre}) and time $t$. The form of the equation is different at $I \le I_m$ and $I \ge I_m$, where $I_m$ is a certain value of the action variable:
\be
\frac{{\partial f}}{{\partial t}} = - V(I) \frac{{\partial f}}{{\partial I}} + \frac12 \frac{{\partial }}{{\partial I}}\left( {D(I) \frac{{\partial f}}{{\partial I}}} \right), \,\,\, \mbox{if} \,\,\, I \le I_m ;
\label{2.1}
\ee
\be
\frac{{\partial f}}{{\partial t}} = - V(I) \frac{{\partial f}}{{\partial I}} - \frac{{\pt V(I) }}{{\pt I}}\left( f - f_*  \right) +  \frac12 \frac{{\partial }}{{\partial I}}\left( {D(I) \frac{{\partial f}}{{\partial I}}} \right), \,\,\, \mbox{if} \,\,\, I \ge I_m .
\label{2.2}
\ee
Here smooth function $V(I)$ has the only maximum at $I=I_m$, is negative at $I_- <I<I_+$ and zero otherwise. Function $D(I)$ is also smooth and positive at $I_l <I<I_r$ and zero otherwise; $I_l<I_-<I_m<I_+<I_r$. Thus, for every value $V_0$ of function $V(I)$ (except for $V(I_m)$) there are two values of $I$ such that $V(I) = V_0$. In the case $I > I_m$, introduce $I_* < I_m$ such that $V(I_*) = V(I)$. Then $f_*$ in (\ref{2.2}) denotes $f(t,I_*)$. Near $I=I_+, \, I=I_-$ function $V(I)$ has the following asymptotics: $V(I) \sim |I-I_{\pm}|^{5/4}$ (see Appendix).

\subsection{Conservation of the number of particles}

The equation (\ref{2.1}-\ref{2.2}) preserves the total number of particles. Indeed, using (\ref{2.1}-\ref{2.2}) and the properties of functions $V(I)$ and $D(I)$ one  obtains
\bea
\frac{\dd}{\dd t} \int_{I_l}^{I_r} f(t,I) \dd I &=& \int_{I_l}^{I_r} \frac{\pt f(t,I)}{\pt t} \dd I \nonumber \\
&=& \int_{I_l}^{I_r} \frac12 \frac{\pt}{\pt I} \left( D\frac{\pt f}{\pt t}\right) \dd I + \int_{I_-}^{I_m}\left(-\frac{\pt}{\pt I}(Vf)+ f\frac{\pt V}{\pt I}\right)\dd I \nonumber \\
&+& \int_{I_m}^{I_+}\left(-\frac{\pt}{\pt I}(Vf)+ f_*\frac{\pt V}{\pt I}\right)\dd I \nonumber \\
&=& \int_{I_-}^{I_m} f(t,I)\frac{\pt V(I)}{\pt I} \dd I + \int_{I_m}^{I_+} f_*(t,\xi)\frac{\pt V(\xi)}{\pt \xi} \dd \xi. \nonumber
\eea
Changing the integration variable in the second integral $\xi \mapsto I$ and using $V(I) = V(\xi), f(t,I) = f_*(t,\xi)$ one finds
\be
\frac{\dd}{\dd t} \int_{I_l}^{I_r} f(t,I) \dd I = \int_{I_-}^{I_m} f(t,I)\frac{\pt V(I)}{\pt I} \dd I + \int_{I_m}^{I_-} f(t,I)\frac{\pt V(I)}{\pt I} \dd I =0. \nonumber
\ee

\subsection{A unique smooth stationary solution: $f = \const$}

Stationary solutions to (\ref{2.1}-\ref{2.2}) meet the following equations:
\bea
- V(I) \frac{{\partial f}}{{\partial I}} + \frac12 \frac{{\partial }}{{\partial I}}\left( {D(I) \frac{{\partial f}}{{\partial I}}} \right) = 0, \,\,\, \mbox{if} \,\,\, I \le I_m; \nonumber \\
- V(I) \frac{{\partial f}}{{\partial I}} - \frac{{\pt V(I) }}{{\pt I}}\left( f - f_*  \right) +  \frac12 \frac{{\partial }}{{\partial I}}\left( {D(I) \frac{{\partial f}}{{\partial I}}} \right) = 0, \,\,\, \mbox{if} \,\,\, I \ge I_m .
\label{2.3}
\eea

In the ranges $I \le I_-$ and $I \ge I_+$ we have $V(I) = 0$. Hence, at these values of $I$ (\ref{2.3}) is just $\frac{{\partial }}{{\partial I}}\left( {D(I) \frac{{\partial f}}{{\partial I}}} \right) = 0$. Therefore $D(I) \frac{{\partial f}}{{\partial I}} = \const$. At $I=I_{l,r}$ function $D(I)$ is zero, hence the constant is zero and $\frac{{\partial f}}{{\partial I}}=0$. Thus, at $I \le I_-$ and $I \ge I_+$ we find $f = \const$. Denote the value of $f$ at $I \le I_-$ as $c_-$.

Consider now the range $I_- \le I \le I_m$. Denote $\frac{{\partial f}}{{\partial I}} = u$. We have equation
$$
-Vu + \frac12 \frac{\dd}{\dd I}(Du) = 0;
$$
its solution is $u = \frac{c}{D} \exp{\int (2V/D) \dd I}$, where $c$ is a constant. As we have already found in the previous paragraph, $u(I_-) = \pt f/\pt I |_{I_-} = 0$. Therefore the constant $c = 0$, and $\pt f/\pt I  = 0$ at $I \in [I_-, I_m]$. Hence, $f \equiv \const = c_-$.

Finally, consider the range $I_m \le I \le I_+$. For values of $I$ in this range, one should put $f_* =c_-$ in (\ref{2.3}). Denote $\tilde f = f - c_-$. Then the second equation in (\ref{2.3}) takes the form
$$
-V \frac{\pt \tilde f}{\pt I} - \frac{ \pt V}{\pt I}\tilde f + \frac12 \frac{\pt}{\pt I}\left( D\frac{\pt \tilde f}{\pt I} \right)=0.
$$
It straightforwardly follows that
\be
-V \tilde f +\frac12 D \frac{\pt \tilde f}{\pt I} = C,
\label{2.4}
\ee
where $C$ is a constant. At $I=I_+$ we have $V(I_+) = 0$ and $ \pt \tilde f/\pt I |_{I_+} = 0$. Therefore, in (\ref{2.4}) the constant $C = 0$, and we have
$$
-V \tilde f +\frac12 D \frac{\pt \tilde f}{\pt I} = 0.
$$
The latter equation has the solution $\tilde f = c_1 \exp{\int (2V/D) \dd I}$, where $c_1$ is a constant. By continuity of function $f$ we have $\tilde f(I_m) = 0$.  Hence $c_1 = 0$, $\tilde f = 0$ and $f = c_-$.

From the continuity of $f$ it follows that $f = c_-$ also at $I \ge I_+$. Thus we have proven that the only smooth solution to (\ref{2.3}) is $f = \const$.

\section{Stability of the constant stationary solution ($D = 0$)}\label{stability}

To demonstrate that the constant stationary solution to (\ref{2.1})-(\ref{2.2}) is stable, we consider the case of zero diffusion: $D(I) \equiv 0$. Doing so we keep in mind that in the systems with trappings into a resonance and scatterings on the resonance one has $D \ll V $ (in dimensionless units; see, e. g., \cite{Artemyev16:pop:letter,Artemyev17:pre}). On the other hand, the diffusive term normally "plays for" more stability. In the next section we numerically demonstrate validity of the analytical results made at $D=0$ in the general case.

Thus we consider equations
\bea
\frac{{\partial f}}{{\partial t}} &=& - V(I) \frac{{\partial f}}{{\partial I}} , \,\,\, \mbox{if} \,\,\, I \le I_m ; \\
\label{3.1}
\frac{{\partial f}}{{\partial t}} &=& - V(I) \frac{{\partial f}}{{\partial I}} - \frac{{\pt V(I) }}{{\pt I}}\left( f - f_*  \right) , \,\,\, \mbox{if} \,\,\, I \ge I_m .
\label{3.2}
\eea
Similarly to Section 2, one can prove that equations (\ref{3.1})-(\ref{3.2}) possess a unique smooth stationary solution $f=\const$. We are going to prove that a smooth solution to (\ref{3.1})-(\ref{3.2}) with initial condition $f(0,I) = f_0(I)$ tends to a constant as $t \to \infty$.

\subsection{General solution}

First, we construct the general solution to equations (\ref{3.1})-(\ref{3.2})  on $I\in(I_-,I_+), \, t\in(0,+\infty)$.

Equations (\ref{3.1})-(\ref{3.2}) can be considered as quasilinear PDEs. To find general solutions to these equations, we consider ODEs for the characteristic curves of (\ref{3.1})-(\ref{3.2}) (see, e. g., \cite{CourantHilbert}):
\bea
\frac{\dd t}{1} &=& \frac{\dd I}{V(I)} = \frac{\dd f}{0}, \,\,\, \mbox{if} \,\,\, I \le I_m ;
\label{3.3} \\
\frac{\dd t}{1} &=& \frac{\dd I}{V(I)} = \frac{\dd f}{-\frac{\pt V}{\pt I}(f-f_*)}, \,\,\, \mbox{if} \,\,\, I \ge I_m .
\label{3.4}
\eea
The first equality in (\ref{3.3}) leads to
$$
t - \int_{I_m}^I \frac{\dd \xi}{V(\xi)} = C_1,
$$
where $C_1 = \const$. Hence, the general solution to (\ref{3.1}) is
\be
f(t,I) = Q\left( t - \int_{I_m}^I \frac{\dd \xi}{V(\xi)}\right), \,\,\, I \le I_m,
\label{3.5}
\ee
where $Q(x)$ is an arbitrary smooth function.

From the first equality in (\ref{3.4}) one finds
$$
t - \int_{I_m}^I \frac{\dd \xi}{V(\xi)} = C_2,
$$
where $C_2 = \const$. The second equality can be rewritten as a linear ODE
\be
\frac{\dd f}{\dd I} = -\frac{1}{V}\frac{\pt V}{\pt I}(f-f_*), \,\,\, I \ge I_m .
\label{3.6}
\ee
The corresponding homogeneous equation
$$
\frac{\dd f}{\dd I} = -\frac{1}{V}\frac{\pt V}{\pt I}f
$$
has a solution $f = C/V$, $C = \const$. To find the general solution to (\ref{3.6}), we put $C = C(I)$ to find
$$
C(I) =  \int_{I_m}^I \frac{\pt V(\xi)}{\pt \xi} f_*\left(C_2 + \int_{I_m}^{\xi}\frac{\dd \eta}{V(\eta)},\xi \right)\dd \xi.
$$
Substituting $C_2 = t-\int_{I_m}^I \dd \eta / V(\eta)$ we find the general solution to (\ref{3.6}):
\bea
f(t,I) &=& \frac{1}{V(I)}\int_{I_m}^I \frac{\pt V(\xi)}{\pt \xi} f_*\left(t + \int_{I}^{\xi}\frac{\dd \eta}{V(\eta)},\xi \right)\dd \xi \nonumber \\
&+& \frac{1}{V(I)}P\left(t - \int_{I_m}^I \frac{\dd \xi}{V(\xi)}\right), \,\,\,  I\ge I_m,
\label{3.7}
\eea
where $P(x)$ is an arbitrary smooth function.

A solution to (\ref{3.1})-(\ref{3.2}) should be continuous at $I = I_m$. Thus, from (\ref{3.5}) and (\ref{3.7}) we find
\be
P(t) = V_m Q(t),
\label{3.8}
\ee
where we introduced the notation $V_m = V(I_m)$.

Consider a value $I_0$ of $I$ such that $I_0\ge I_m$. Introduce function $\La (I) \le I_m$ such that its values satisfy $V(\La(I_0)) = V(I_o)$. (In terms of trapping into the resonance it means that a particle trapped at $I=\La(I_0)$ escapes from the resonance at $I = I_0$, see \cite{Artemyev16:pop:letter,Artemyev17:pre}.) Using this notation one can write
\be
f_*(t,I) = f(t, \La(I)) = Q \left(t- \int_{I_m}^{\La(I)} \frac{\dd \xi}{V(\xi)} \right).
\label{3.9}
\ee
Now, from (\ref{3.7}), (\ref{3.8}), and (\ref{3.9}) it is straightforward to obtain
\bea
f(t,I) &=& \frac{1}{V(I)}\int_{I_m}^I \frac{\pt V(\xi)}{\pt \xi} Q\left(t - \int_{I_m}^{I}\frac{\dd \eta}{V(\eta)}- \int_{\xi}^{\La(\xi)}\frac{\dd \eta}{V(\eta)} \right)\dd \xi \nonumber \\
&+& \frac{V_m}{V(I)}Q\left(t - \int_{I_m}^I \frac{\dd \xi}{V(\xi)}\right), \,\,\,  I\ge I_m.
\label{3.10}
\eea

\subsection{Solution with given initial conditions (Cauchy problem)}

Introduce notation $f_0(I) = f(0,I)$. From (\ref{3.5}) and (\ref{3.10}) at $t=0$ we have
\be
f_0(I) = Q\left( - \int_{I_m}^I \frac{\dd \xi}{V(\xi)}\right), \,\,\, I \le I_m,
\label{3.11}
\ee
\bea
f_0(I) &=& \frac{1}{V(I)}\int_{I_m}^I \frac{\pt V(\xi)}{\pt \xi} Q\left( - \int_{I_m}^{I}\frac{\dd \eta}{V(\eta)}- \int_{\xi}^{\La(\xi)}\frac{\dd \eta}{V(\eta)} \right)\dd \xi \nonumber \\
&+& \frac{V_m}{V(I)}Q\left(- \int_{I_m}^I \frac{\dd \xi}{V(\xi)}\right), \,\,\,  I\ge I_m.
\label{3.12}
\eea
Now, we are to prove that formulas (\ref{3.5}), (\ref{3.10}) together with initial conditions (\ref{3.11}), (\ref{3.12}) indeed provide a bounded and smooth solution to equations (\ref{3.1}), (\ref{3.2}). In order to do this, we prove that there exists a unique smooth bounded function $Q$ satisfying (\ref{3.11}), (\ref{3.12}).

Note that $V(I)\le 0$. Hence $\left(- \int_{I_m}^I \dd \xi/V(\xi)\right) \le 0$ at $I \le I_m$, and $\left(- \int_{I_m}^I \dd \xi/V(\xi)\right)\ge 0$ at $I \ge I_m$. At values of $I$ close to $I_{\pm}$ we have $V(I) \sim |I-I_{\pm}|^{5/4}$ (see Appendix) and hence the integral $\int_{I_m}^{I_{\pm}} \dd \xi/V(\xi)$ diverges. Therefore, given $f_0(I)$, equation (\ref{3.11}) defines $Q(x)$ at all $x\in(-\infty,0)$. Moreover, as $f_0(I)$ is bounded, (\ref{3.11}) implies that $Q(x) \to f_0(I_-)$ as $x \to -\infty$.

At $I \ge I_m$, function $Q(x)$ is defined by (\ref{3.12}). Introduce new integration variable $x$ instead of $\xi$:
\be
x = - \int_{I_m}^{I}\frac{\dd \eta}{V(\eta)}- \int_{\xi}^{\La(\xi)}\frac{\dd \eta}{V(\eta)}.
\label{A.2}
\ee
One has
\be
\dd x = \left( -\frac{1}{V(\La(\xi))}\frac{\pt\La}{\pt\xi} + \frac{1}{V(\xi)} \right) \dd \xi.
\label{A.3}
\ee
By its definition, function $\La(\xi)$ satisfies $V(\xi) = V(\La(\xi))$. Differentiating the latter equality one obtains
$$
\frac{\pt\La}{\pt\xi} = \frac{V^{\prime}(\xi)}{V^{\prime}(\La(\xi))},
$$
where the prime denotes derivative of the function over its argument. Substituting into (\ref{A.3}), one obtains
\be
\frac{\pt V}{\pt \xi} = g(I,x) \dd x,
\label{A.4}
\ee
where we have introduced the notation
\be
g(I,x) = \frac{1}{(V(\xi)V^{\prime}(\xi))^{-1} - (V(\La(\xi))V^{\prime}(\La(\xi)))^{-1}}.
\label{A.5}
\ee
Equation (\ref{3.12}) takes the form:
\be
V(I)f_0(I) = \int_{a(I)}^{b(I)} g(I,x)Q(x) \dd x + V_m Q\left(- \int_{I_m}^I \frac{\dd \xi}{V(\xi)}\right),\,\,\,  I\ge I_m,
\label{A.6}
\ee
where
\bea
a(I) = -\int_{I_m}^{I}\frac{\dd \eta}{V(\eta)} >0,
\label{A.7} \\
b(I) = - \int_{I_m}^{\La(I)}\frac{\dd \eta}{V(\eta)} <0.
\label{A.8}
\eea

At $x \le 0$ function $Q(x)$ is defined in terms $f_0(I), \, I \le I_m$ (see (\ref{3.11})). Therefore, the integral $\int_0^{b(I)} g(I,x)Q(x) \dd x$ is a known function of $I$, which we denote as $h(I)$. Hence,
\be
-h(I) + V(I)f_0(I) = -\int_0^{a(I)} g(I,x)Q(x) \dd x + V_m Q\left(- \int_{I_m}^I \frac{\dd \xi}{V(\xi)}\right),\,\,\,  I\ge I_m.
\label{A.9}
\ee
It follows from (\ref{A.8}-\ref{A.8}) that $a(I) \to +\infty$ and $b(I) \to -\infty$ as $I \to I_+$. Using definitions of functions $g(I,x)$ and $h(I)$ and returning to the original integration variable $\xi$ one can see that $h(I)$ is bounded as $I \to I_+$. Moreover, the integral $\int_0^{a(I)}g(I,x)\dd x$ converges as $I \to I_+$.

Equation (\ref{A.9}) is a Volterra integral equation of the second kind with $a(I)$ considered as an independent variable (see, e. g.,  \cite{CourantHilbert}). On any bounded segment $(0,A)$ of values of $a$ this equation has a unique continuous solution. Therefore, it has a unique continuous solution on any segment $(I_m,I_k)$ of values of $I$, where $I_k <I_+$.

Rewrite (\ref{A.9}) as
\bea
-h(I) + V(I)f_0(I) &=& -\int_0^A g(I,x)Q(x) \dd x -\int_A^{a(I)} g(I,x)Q(x) \dd x \nonumber \\
&+& V_m Q\left(- \int_{I_m}^I \frac{\dd \xi}{V(\xi)}\right),\,\,\,  I\ge I_m.
\label{A.10}
\eea
In the first integral in (\ref{A.10}) $Q(x)$ is a known bounded function. Hence, this integral converges for any finite value of $A$ (i.e., for any $I < I_+$). Thus, we can rewrite the equation in the form
\be
-\tilde h(I)  =  - \frac{1}{V_m} \int_A^{a(I)} g(I,x)Q(x) \dd x +  Q\left(- \int_{I_m}^I \frac{\dd \xi}{V(\xi)}\right),
\label{A.11}
\ee
where $\tilde h(I)$ is a known function and we consider such values of $I$ that $a(I) > A$. The integral $\int_A^{a(I)}g(I,x)\dd x$ converges (see above) and is small at large enough values of $A$. Therefore, operator
$$
Q(a(I)) - \frac{1}{V_m} \int_A^{a(I)} g(I,x)Q(x) \dd x
$$
is a contraction (see, e. g., \cite{CourantHilbert}). Hence, the equation has a unique continuous bounded solution. Smoothness of $Q(x)$ at $x<0$ is due to smoothness of the initial condition $f_0(I)$ at $I < I_m$. Smoothness of $Q(x)$ at $x>0$ follows from its boundedness and equation (\ref{A.6}). Thus, the constructed solution $f(t,I)$ is smooth at $I < I_m$ and $I > I_m$. To prove smoothness of $f(t,I)$ at $I=I_m$ one can differentiate (\ref{3.5}) and (\ref{3.10}) with respect to $I$ and find the values of the both derivatives at $I \to I_m$. Taking into account that $\pt V/\pt I = 0$ at $I = I_m$ it is straightforward to obtain that the both values of $\pt f/\pt I$ coincide.

\subsection{Behavior at $t \to \infty$}

Now we can study behavior of the constructed solution $f(t,I)$ as $t \to +\infty$.

According to the previous subsection, function $Q(x)$ is bounded. Therefore, a finite limit exists:
\be
W = \lim_{I \to I_+} \int_{I_m}^I \frac{\pt V(\xi)}{\pt \xi} Q\left( - \int_{I_m}^{I}\frac{\dd \eta}{V(\eta)}- \int_{\xi}^{\La(\xi)}\frac{\dd \eta}{V(\eta)} \right)\dd \xi .
\label{3.13}
\ee
Indeed, $\pt V/\pt\xi = O(|\xi-I_{\pm}|^{1/4})$ if $\xi$ is close to $I_{\pm}$, and $\pt V/\pt\xi = O(|\xi-I_m|)$ if $\xi$ is close to $I_m$. If the argument of $Q$ in (\ref{3.13}) is negative, $Q$ is a known (in terms of $f_0(I)$, see (\ref{3.11})) bounded function. Taking into account that $V(I) \to 0$ as $I \to I_+$ and $f_0(I)$ is bounded, we find from (\ref{3.12}) and (\ref{3.13}) that $Q(x) \to -W/V_m$ as $x \to +\infty$.

Consider now the limit $\lim_{t\to\infty} f(t,I)$ at a fixed $I \ne I_{\pm}$. At $I \le I_m$ we have
\be
f(t,I) = Q\left( t - \int_{I_m}^I \frac{\dd \xi}{V(\xi)}\right) \to -\frac{W}{V_m} = \const, \,\,\, \mbox{as} \,\, t \to \infty.
\label{3.14}
\ee
Hence, for $I \ge I_m$ one has
\be
f_*(t,I)\to -\frac{W}{V_m} = \const, \,\,\, \mbox{as} \,\, t \to \infty.
\label{3.15}
\ee
Therefore, at $I \ge I_m$, we find from (\ref{3.7}), (\ref{3.8}) that
\be
f(t,I) \to \frac{1}{V(I)} \left( -\frac{W}{V_m} \int_{I_m}^I \frac{\pt V(\xi)}{\pt \xi} \dd\xi - W \right) = -\frac{W}{V_m}, \,\,\, \mbox{as} \,\, t \to \infty.
\label{3.16}
\ee
Thus, we have proved that $f(t,I) \to \const$ as $t \to \infty$. Therefore, the constant solution to equations (\ref{3.1})-(\ref{3.2}) is stable. In the following section we demonstrate that this is also valid for the kinetic equation (\ref{2.1})-(\ref{2.2}) with $D(I) \ne 0$.

\section{Numerical tests}
To verify our analytical results and confirm the main conclusion of stability of the stationary solution $f=\const$, we solve Eqs. (\ref{2.1}, \ref{2.2}) numerically for a set of  initial conditions, $f_0(I)$. Coefficients of Eqs. (\ref{2.1}, \ref{2.2}) and dependence $\Lambda(I)$ are defined from the analysis of Hamiltonian (\ref{A1.1}) (see Appendix). We use a sample function $V(I)=- \varepsilon^{1/2}(1-(I/I_a)^2)^{5/4}$ with $I_{\pm}=\pm I_a$ and $I_m=0$ (thus, $\Lambda(I)=-I$) with $\varepsilon=10^{-3}$ (we put $V(I)=0$ for $|I|>I_a$). The diffusion coefficient is chosen in the form $D=D_0\varepsilon(1-I^2)$ and differs from zero at $I\in[-1,1]$, i.e. if $I_a<1$ the $I$-range of nonlinear drift and trapping is shorter than the range of diffusion. The initial distribution is $f_0(I)=C_0\exp(-(I-I_0)^2/\delta I^2)$ where $\delta I=1/2$, $C_0$ is defined by normalization $\int_{-1}^{+1}{f_0(I)\dd I}=1$, and $I_0=0,\pm 1/2$ (this distribution is slightly modified around $I=I_{\pm}$ to satisfy the boundary conditions $\dd f/\dd I|_{I_{\pm}}=0$). This set of parameters assumes that particles (with $f(I,t)$ distribution) drift toward smaller $I$ and are transported (due to trapping) from negative $I$ to positive $I$. Therefore, the initial distribution with $I_0=-1/2$ peaks within the region where charged particles can be trapped ($\dd V/\dd I<0$). The evolution of this distribution includes both drift and transport from the very beginning. The initial distribution with $I_0=1/2$ peaks within the region where trapping is impossible ($\dd V/\dd I>0$). This distribution should first drift to the region of $I<0$ (where $\dd V/\dd I<0$), and only then the transport due to trapping can start. The initial distribution with $I_0=0$ peaks at $\dd V/\dd I=0$. ] The evolution of this distribution should resemble spreading, because the distribution  drifts to smaller $I$ (larger $|I|$) and at the same time some portion of these particles (with $I<0$) is transported toward large $I$.

\begin{figure*}
\centering
\includegraphics[width=0.95\textwidth]{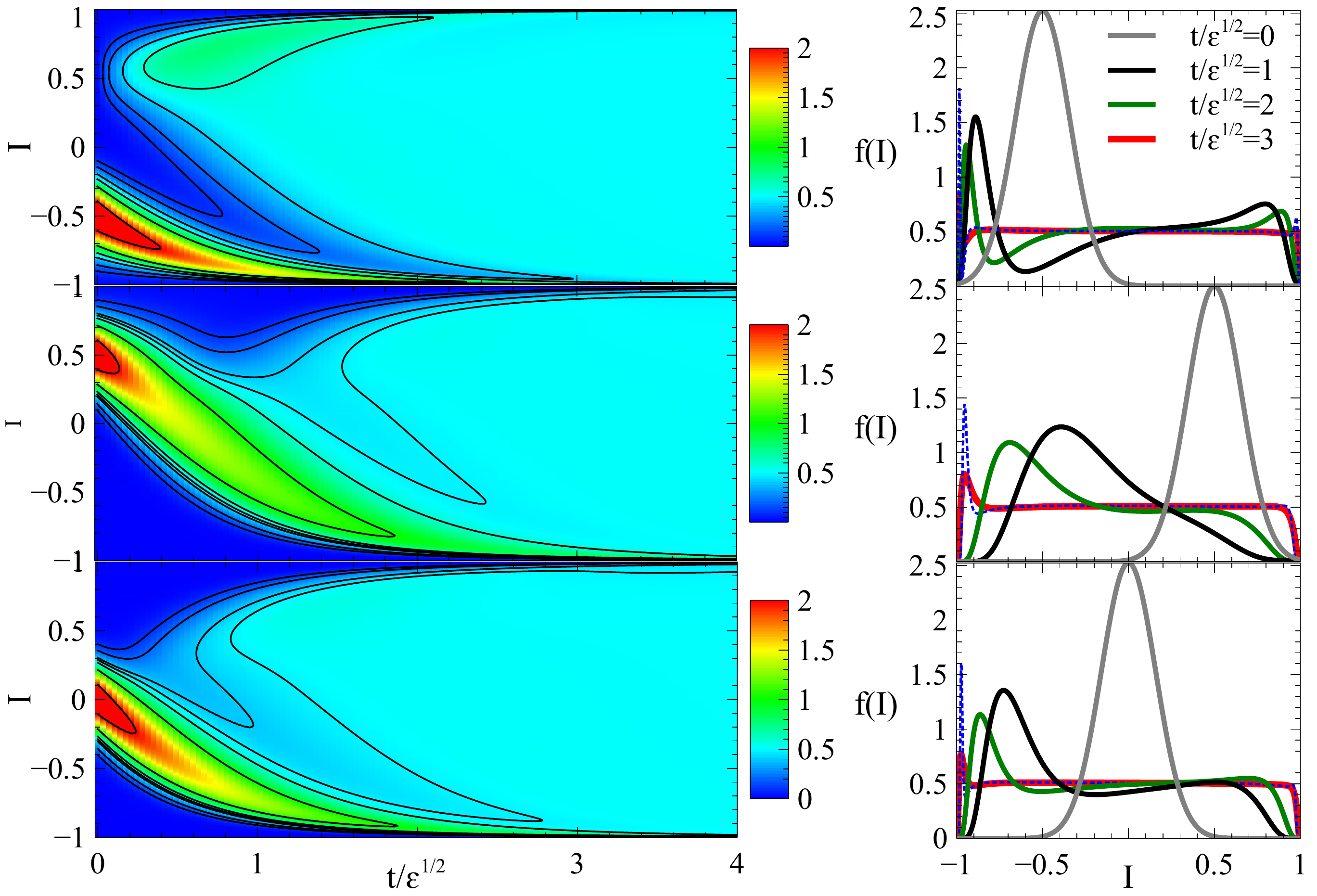}
  \caption{Solutions of Eqs. (\ref{2.1}, \ref{2.2}) with three initial distributions $f_0(I)=C_0\exp(-(I-I_0)^2/\delta I^2)$ with $I_0=-1/2$ (top panels), with $I_0=1/2$ (middle panels), and with $I_0=0$ (bottom panels). Left panels show 2D distributions $f(t,I)$ and right panels show $f(t,I)$ at four time moments. Blue dashed lines show $f(t,I)$ at $t/\sqrt{\varepsilon}=3$ for the solution with $D_0=1/10$ (all other curves and 2D distributions are  for solutions with $D_0=1$). The boundary conditions are $\dd f/\dd I=0$ at $I=\pm 1$, and the drift velocity is $V(I)=- \varepsilon^{1/2}(1-(I/I_a)^2)^{5/4}$ with $I_a=1$, i.e. the diffusion and nonlinear drift (and trapping) are acting on the same $I$ range.
\label{fig:distributions1}}
\end{figure*}

\begin{figure*}
\centering
\includegraphics[width=0.95\textwidth]{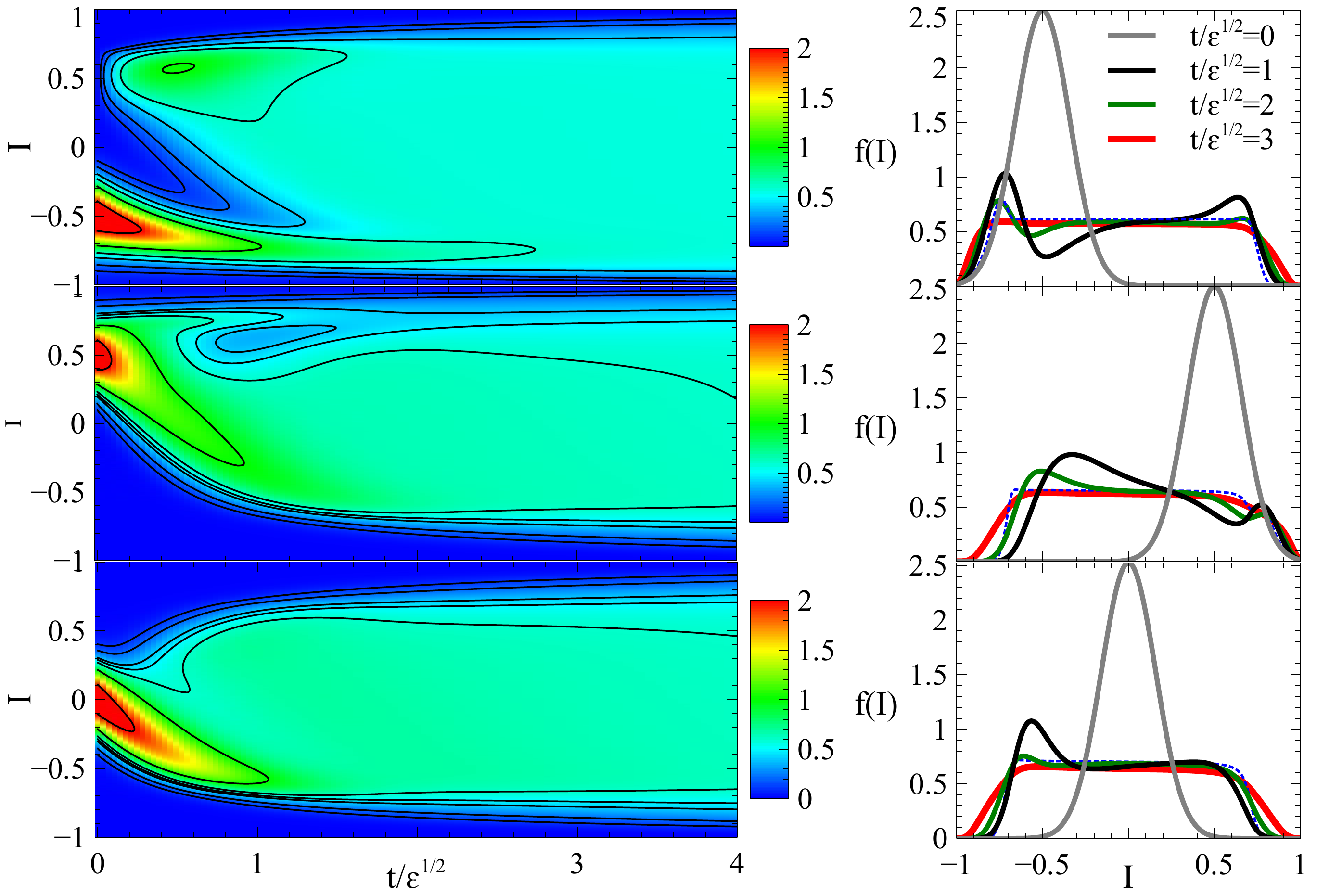}
  \caption{Solutions of Eqs. (\ref{2.1}, \ref{2.2}) with three initial distributions $f_0(I)=C_0\exp(-(I-I_0)^2/\delta I^2)$ with $I_0=-1/2$ (top panels), with $I_0=1/2$ (middle panels), and with $I_0=0$ (bottom panels). Left panels show 2D distributions $f(t,I)$ and right panels show $f(t,I)$ at four time moments. Blue dashed lines show $f(t,I)$ at $t/\sqrt{\varepsilon}=3$ for the solution with $D_0=1/10$ (all other curves and 2D distributions are  for solutions with $D_0=1$).  The boundary conditions are $\dd f/\dd I=0$ at $I=\pm 1$, and the drift velocity  is $V(I)=- \varepsilon^{1/2}(1-(I/I_a)^2)^{5/4}$ with $I_a=1/\sqrt{2}$, i.e. the nonlinear drift and trappings  act on a shorter $I$-range than the diffusion.
\label{fig:distributions2}}
\end{figure*}

We consider two sets of solutions: with $I_a=1$ (Fig. \ref{fig:distributions1}) and with $I_a=1/\sqrt{2}$ (Fig. \ref{fig:distributions2}). In the first set of solutions, the distribution $f(t,I)$ changes slowly  near the boundary $I_{\pm}$ because both $V$ and $D$ vanish there, but  eventually this evolution form the flat distribution $f\sim \const$. In the second set of solutions, the boundary of the $I$-range of nonlinear drift and trappings is located at $I=\pm I_a$ where the diffusion is still quite strong (not vanishing). This results in the absence of peculiarities of particle distribution at this boundary: the nonlinear processes (drift and trapping) rapidly form $f\sim \const$ distribution at $|I|<I_a$ and then slow diffusion moves boundaries of this distribution toward the system boundary, $I=\pm 1$.

Figures \ref{fig:distributions1}, \ref{fig:distributions2}  show results of numerical solutions of Eqs. (\ref{2.1}, \ref{2.2}). Left panels demonstrate 2D distributions $f(t,I)$, and right panels show $f(t,I)$ at four time moments. The initial distribution determines the beginning of  the evolution, but after $t/\sqrt{\varepsilon}\sim 2$ distribution $f(t,I)$ becomes quite flat in all three examples. The diffusion contributes to $f$ spreading, but does not change the general evolution which is determined by drift and trapping effects (compare red and blue curves at the right panels; these curves are plotted for solutions with $D_0=1$ and $D_0=1/10$). Thus, the numerical solution confirms our main conclusion: independently on initial distribution $f_0(I)$ the solution of Eqs. (\ref{2.1}, \ref{2.2}) tends to the uniform one $f=\const$.

\section{Discussion and conclusions}
\begin{figure*}
\centering
\includegraphics[width=0.55\textwidth]{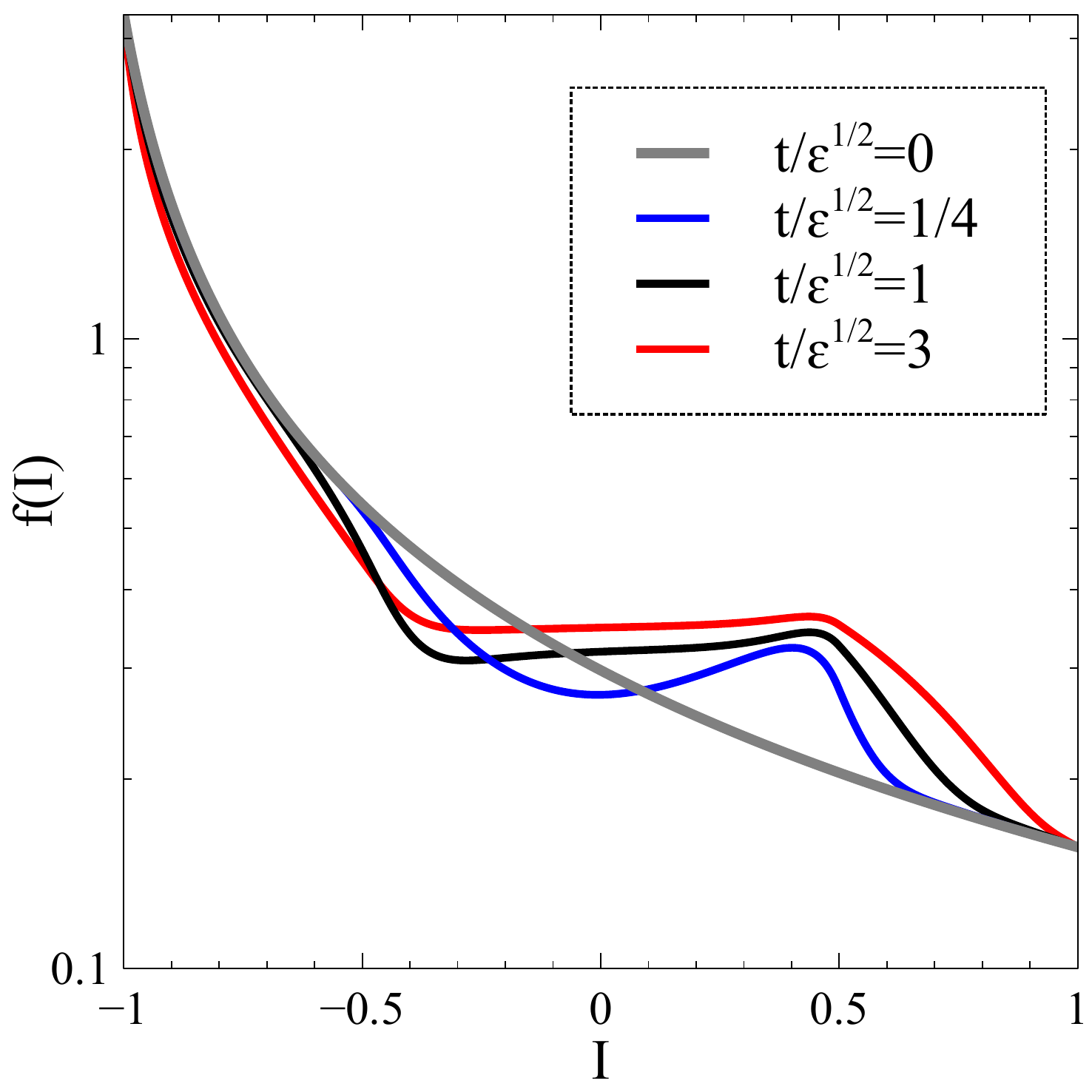}
\caption{Solutions of Eqs. (\ref{2.1}, \ref{2.2}) for the initial distribution $f_0(I)=C_0/(I_0+I)$ with $I_0=11/10$. The boundary conditions are $df/dI=0$ at $I=\pm 1$, and the drift velocity equation is $V(I)=- \varepsilon^{1/2}(1-(I/I_a)^2)^{5/4}$ with $I_a=1/2$, i.e. the nonlinear drift and trappings act on a shorter $I$-range than the diffusion.
\label{fig:distributions3}}
\end{figure*}

In this paper we show that the kinetic equation including nonlinear effects of particle trapping and scattering by high-amplitude waves has a unique stationary solution, $f = \const$, and all solutions with smooth initial conditions tend to constant as $t\to \infty$. Thus, despite the significant difference between this equation (i.e., Eqs. (\ref{2.1}, \ref{2.2})) and the classical diffusion equation describing the quasi-linear wave-particle interaction \cite[e.g.,][]{Drummond&Pines62,Vedenov62}, for  the both  equations the solutions tend to constant. The main difference between nonlinear resonant interaction including trappings and drift and the quasi-linear diffusion is the time-scale required   to form a constant solution. The typical time-scale of the nonlinear processes is $\sim 1/V\sim \varepsilon^{-1/2}$, whereas the time-scale of diffusion varies from $1/D\sim 1/\varepsilon$ \cite[e.g.,][]{Karpman&Shkliar77} to $1/D\sim 1/\varepsilon^2$ \cite{Kennel&Engelmann66} depending on the characteristics of a particular plasma system. Thus, in the $I$-range where nonlinear wave-particle interaction is possible, the solution $f=\const$ is formed much faster than in the $I$-range where only diffusion affects the particle distribution (see, e.g., Fig. \ref{fig:distributions2}). In the realistic plasma systems where a value $(I+\const)$ plays the role of the particle energy \cite[see, e.g.,][]{Shklyar81, Albert93, Artemyev17:pre}, distributions $f(t,I)$ have a typical shape of energy spectra, i.e. $f$ decreases as $I$ increase. For such distributions the nonlinear wave-particle interaction within a limited $I$-range results in formation of a plateau and following evolution of this plateau due to diffusion (see, for example, Fig. \ref{fig:distributions3}). Such types of distributions with energy-limited plateau are usually observed in the near-Earth plasma systems where nonlinear wave-particle interaction is supposed to be strong \cite[e.g.,][]{Min14, Li17:Langmuir, Kurita18}. Note also that such regions of flattened distribution might be of interest in problems of tailoring barriers in the phase space in laboratory plasmas (c. f., e. g., \cite{OgawaLeonciniVasilievGarbet_preprint}).

In conclusion, we have shown that the kinetic equation including effects of nonlinear wave-particle resonant interaction (scattering-induced drift and trappings) has a unique stationary solution $f=\const$. This solution is stable and any solution $f(t,I)$ with smooth initial conditions tends to constant as $t\to\infty$. Using numerical simulations, we have checked that particle diffusion does not influence significantly these conclusions. Obtained results could be useful for analysis of various plasma systems where nonlinear wave-particle interaction plays an important role.

\section*{Appendix}

\begin{figure*}
\centering
\includegraphics[width=0.95\textwidth]{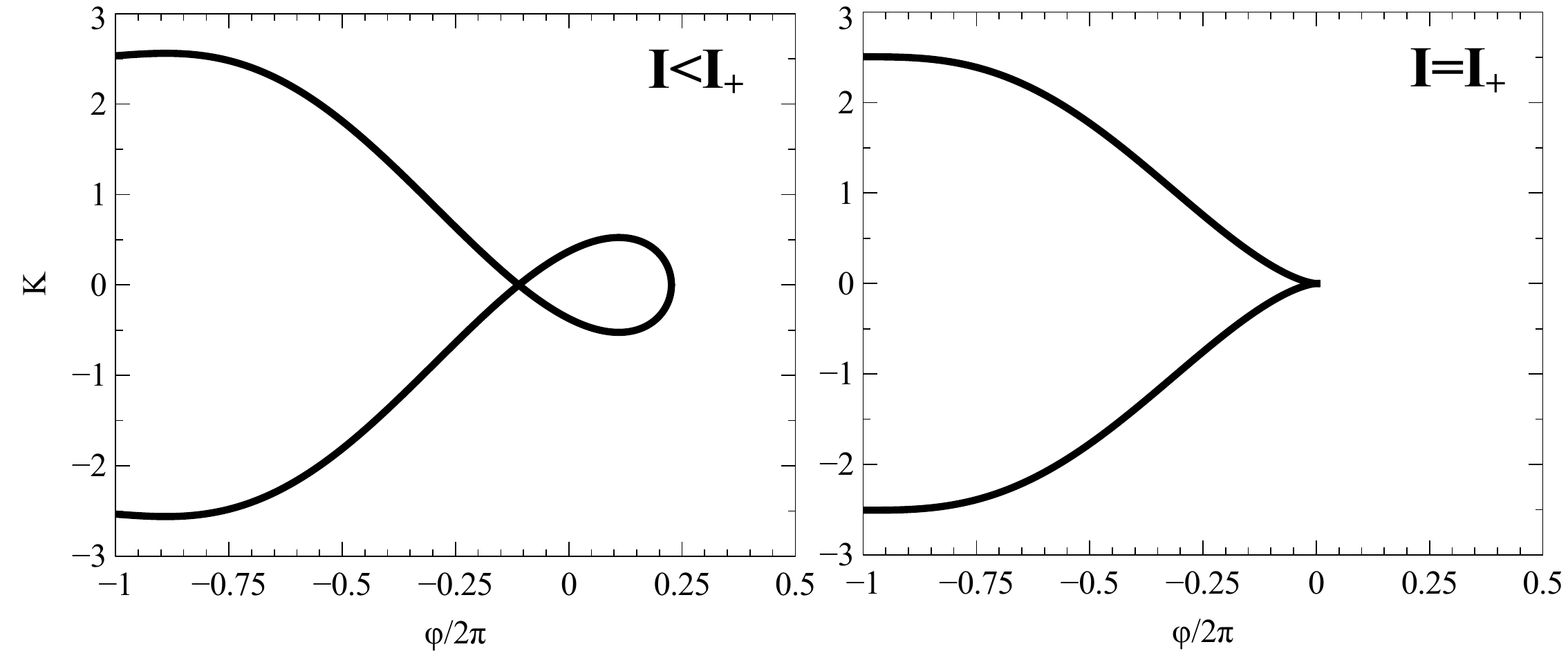}
  \caption{Phase portraits of system (\ref{A1.1}) for $I<I_+$ (left panel) and $I=I_+$ (right panel).
\label{fig:phaseportrait}}
\end{figure*}

Function $V(I)$ is proportional to the area bounded by the separatrix loop on the phase portrait of the system near the resonance (see \cite{Artemyev16:pop:letter,Artemyev17:pre,Artemyev17:arXiv}). The corresponding Hamiltonian (called "pendulum-like" Hamiltonian, see, e. g., \cite{Neishtadt99,Neishtadt06}) has the form
\be
F = \frac12 g(I)K^2 - A(I)\sin\ffi + \beta(I)\ffi,
\label{A1.1}
\ee
where $K,\ffi$ are canonically conjugate variables and $I$ can be considered as a parameter; $g,A,\beta$ are functions of $I$; one may assume that they are positive. (Note that in \cite{Artemyev17:arXiv} the variable corresponding to our $I$ is denoted as $J$). If $A(I)>\beta(I)$, there is the separatrix on the phase portrait. At $I=I_{\pm}$ one has $A=\beta$, and the saddle-center bifurcation occurs. Consider for definiteness $I=I_+$. At $I=I_+$, the separatrix loop disappears at $\ffi = 0$ (see Fig \ref{fig:phaseportrait}).  Expanding Hamiltonian $F$ near  $I=I_+,\ffi=0$ one obtains in the main approximation
\be
F = \frac12 g(I_+) K^2 - \dt \ffi +\tilde A \ffi^3,
\label{A1.2}
\ee
where $\dt = (A^{\prime}(I_+)-\beta^{\prime}(I_+))(I - I_+)$; $\tilde A = A(I_+)/6$. The area inside the separatrix loop on the phase portrait (at $\dt>0$) is
\be
S(\dt) = 2 \int_{\ffi_1}^{\ffi_2} K \dd \ffi = 2 \int_{\ffi_1}^{\ffi_2} \sqrt{\frac{2}{g}(F_s+\dt \ffi - \tilde A \ffi^3)} \dd \ffi,
\label{A1.3}
\ee
where $\ffi_1,\ffi_2 \sim \sqrt{\dt}$ are the zeros of the integrand, and $F_s \sim \dt^{3/2}$ is the value of the Hamiltonian $F$ at the saddle point. At $\dt \ll 1$, one  obtains $S(\dt) \sim \dt^{5/4}$. Therefore, $V(I)$ at $I$ close to $I_{\pm}$ has the asymptotics $V\sim |I-I_{\pm}|^{5/4}$.

\section*{Acknowledgements}
The work was supported by
the Russian Scientific Fund, Project No. 14-12-00824.

\newpage
\section*{References}

\begin{thebibliography}{10}
\expandafter\ifx\csname url\endcsname\relax
  \def\url#1{\texttt{#1}}\fi
\expandafter\ifx\csname urlprefix\endcsname\relax\def\urlprefix{URL }\fi
\expandafter\ifx\csname href\endcsname\relax
  \def\href#1#2{#2} \def\path#1{#1}\fi

\bibitem{Spitzer62}
L.~{Spitzer}, {Physics of Fully Ionized Gases}, 1962.

\bibitem{Galeev&Sagdeev79}
A.~A. {Galeev}, R.~Z. {Sagdeev}, {Nonlinear Plasma Theory}, in: A.~M.~A.
  {Leontovich} (Ed.), Reviews of Plasma Physics, Volume 7, Vol.~7 of Reviews of
  Plasma Physics, 1979, p.~1.

\bibitem{Gary:book05}
S.~P. Gary, The Theory of Plasma Waves, Cambridge Atmospheric and Space,
  Cambridge University Press, 2005.

\bibitem{Krasnoselskikh13}
V.~{Krasnoselskikh}, M.~{Balikhin}, S.~N. {Walker}, S.~{Schwartz},
  D.~{Sundkvist}, V.~{Lobzin}, M.~{Gedalin}, S.~D. {Bale}, F.~{Mozer},
  J.~{Soucek}, Y.~{Hobara}, H.~{Comisel}, {The Dynamic Quasiperpendicular
  Shock: Cluster Discoveries}, \ssr 178 (2013) 535--598.
\newblock \href {http://arxiv.org/abs/1303.0190} {\path{arXiv:1303.0190}},
  \href {http://dx.doi.org/10.1007/s11214-013-9972-y}
  {\path{doi:10.1007/s11214-013-9972-y}}.

\bibitem{Wing14}
S.~{Wing}, J.~R. {Johnson}, C.~C. {Chaston}, M.~{Echim}, C.~P. {Escoubet},
  B.~{Lavraud}, C.~{Lemon}, K.~{Nykyri}, A.~{Otto}, J.~{Raeder}, C.-P. {Wang},
  {Review of Solar Wind Entry into and Transport Within the Plasma Sheet}, \ssr
  184 (2014) 33--86.
\newblock \href {http://dx.doi.org/10.1007/s11214-014-0108-9}
  {\path{doi:10.1007/s11214-014-0108-9}}.

\bibitem{Fujimoto11}
M.~{Fujimoto}, I.~{Shinohara}, H.~{Kojima}, {Reconnection and Waves: A Review
  with a Perspective}, \ssr 160 (2011) 123--143.
\newblock \href {http://dx.doi.org/10.1007/s11214-011-9807-7}
  {\path{doi:10.1007/s11214-011-9807-7}}.

\bibitem{Lysak90}
R.~L. {Lysak}, {Electrodynamic coupling of the magnetosphere and ionosphere},
  \ssr 52 (1990) 33--87.
\newblock \href {http://dx.doi.org/10.1007/BF00704239}
  {\path{doi:10.1007/BF00704239}}.

\bibitem{Thorne10:GRL}
R.~M. {Thorne}, {Radiation belt dynamics: The importance of wave-particle
  interactions}, \grl 372 (2010) 22107.
\newblock \href {http://dx.doi.org/10.1029/2010GL044990}
  {\path{doi:10.1029/2010GL044990}}.

\bibitem{Ni16:ssr}
B.~{Ni}, R.~M. {Thorne}, X.~{Zhang}, J.~{Bortnik}, Z.~{Pu}, L.~{Xie}, Z.-j.
  {Hu}, D.~{Han}, R.~{Shi}, C.~{Zhou}, X.~{Gu}, {Origins of the Earth's Diffuse
  Auroral Precipitation}, \ssr\href
  {http://dx.doi.org/10.1007/s11214-016-0234-7}
  {\path{doi:10.1007/s11214-016-0234-7}}.

\bibitem{Watt&Rankin12}
C.~E.~J. {Watt}, R.~{Rankin}, {Alfv{\'e}n Wave Acceleration of Auroral
  Electrons in Warm Magnetospheric Plasma}, Washington DC American Geophysical
  Union Geophysical Monograph Series 197 (2012) 251--260.
\newblock \href {http://dx.doi.org/10.1029/2011GM001171}
  {\path{doi:10.1029/2011GM001171}}.

\bibitem{Drummond&Pines62}
W.~E. {Drummond}, D.~{Pines}, {Nonlinear stability of plasma oscillations},
  Nuclear Fusion Suppl. 3 (1962) 1049--1058.

\bibitem{Vedenov62}
A.~A. {Vedenov}, E.~{Velikhov}, R.~{Sagdeev}, {Quasilinear theory of plasma
  oscillations}, Nuclear Fusion Suppl. 2 (1962) 465--475.

\bibitem{Matsumoto94}
H.~{Matsumoto}, H.~{Kojima}, T.~{Miyatake}, Y.~{Omura}, M.~{Okada},
  I.~{Nagano}, M.~{Tsutsui}, {Electrotastic Solitary Waves (ESW) in the
  magnetotail: BEN wave forms observed by GEOTAIL}, \grl 21 (1994) 2915--2918.
\newblock \href {http://dx.doi.org/10.1029/94GL01284}
  {\path{doi:10.1029/94GL01284}}.

\bibitem{Panov06}
E.~V. {Panov}, J.~{B{\"u}chner}, M.~{Fr{\"a}nz}, A.~{Korth}, S.~P. {Savin},
  K.-H. {Forna{\c c}on}, I.~{Dandouras}, H.~{R{\`e}me}, {CLUSTER observation of
  collisionless transport at the magnetopause}, \grl 33 (2006) 15109.
\newblock \href {http://dx.doi.org/10.1029/2006GL026556}
  {\path{doi:10.1029/2006GL026556}}.

\bibitem{Bale&Mozer07}
S.~D. {Bale}, F.~S. {Mozer}, {Measurement of Large Parallel and Perpendicular
  Electric Fields on Electron Spatial Scales in the Terrestrial Bow Shock},
  Physical Review Letters 98~(20) (2007) 205001.
\newblock \href {http://arxiv.org/abs/physics/0703101}
  {\path{arXiv:physics/0703101}}, \href
  {http://dx.doi.org/10.1103/PhysRevLett.98.205001}
  {\path{doi:10.1103/PhysRevLett.98.205001}}.

\bibitem{Cattell08}
C.~{Cattell}, J.~R. {Wygant}, K.~{Goetz}, K.~{Kersten}, P.~J. {Kellogg},
  T.~{von Rosenvinge}, S.~D. {Bale}, I.~{Roth}, M.~{Temerin}, M.~K. {Hudson},
  R.~A. {Mewaldt}, M.~{Wiedenbeck}, M.~{Maksimovic}, R.~{Ergun}, M.~{Acuna},
  C.~T. {Russell}, {Discovery of very large amplitude whistler-mode waves in
  Earth's radiation belts}, \grl 35 (2008) 1105.
\newblock \href {http://dx.doi.org/10.1029/2007GL032009}
  {\path{doi:10.1029/2007GL032009}}.

\bibitem{Cully08}
C.~M. {Cully}, J.~W. {Bonnell}, R.~E. {Ergun}, {THEMIS observations of
  long-lived regions of large-amplitude whistler waves in the inner
  magnetosphere}, \grl 35 (2008) 17.
\newblock \href {http://dx.doi.org/10.1029/2008GL033643}
  {\path{doi:10.1029/2008GL033643}}.

\bibitem{Khotyaintsev10}
Y.~V. {Khotyaintsev}, A.~{Vaivads}, M.~{Andr{\'e}}, M.~{Fujimoto},
  A.~{Retin{\`o}}, C.~J. {Owen}, {Observations of Slow Electron Holes at a
  Magnetic Reconnection Site}, Physical Review Letters 105~(16) (2010) 165002.
\newblock \href {http://dx.doi.org/10.1103/PhysRevLett.105.165002}
  {\path{doi:10.1103/PhysRevLett.105.165002}}.

\bibitem{Chaston13}
C.~C. {Chaston}, Y.~{Yao}, N.~{Lin}, C.~{Salem}, G.~{Ueno}, {Ion heating by
  broadband electromagnetic waves in the magnetosheath and across the
  magnetopause}, \jgr 118 (2013) 5579--5591.
\newblock \href {http://dx.doi.org/10.1002/jgra.50506}
  {\path{doi:10.1002/jgra.50506}}.

\bibitem{Wilder16}
F.~D. {Wilder}, R.~E. {Ergun}, K.~A. {Goodrich}, M.~V. {Goldman}, D.~L.
  {Newman}, D.~M. {Malaspina}, A.~N. {Jaynes}, S.~J. {Schwartz}, K.~J.
  {Trattner}, J.~L. {Burch}, M.~R. {Argall}, R.~B. {Torbert}, P.-A.
  {Lindqvist}, G.~{Marklund}, O.~{Le Contel}, L.~{Mirioni}, Y.~V.
  {Khotyaintsev}, R.~J. {Strangeway}, C.~T. {Russell}, C.~J. {Pollock}, B.~L.
  {Giles}, F.~{Plaschke}, W.~{Magnes}, S.~{Eriksson}, J.~E. {Stawarz}, A.~P.
  {Sturner}, J.~C. {Holmes}, {Observations of whistler mode waves with
  nonlinear parallel electric fields near the dayside magnetic reconnection
  separatrix by the Magnetospheric Multiscale mission}, \grl 43 (2016)
  5909--5917.
\newblock \href {http://dx.doi.org/10.1002/2016GL069473}
  {\path{doi:10.1002/2016GL069473}}.

\bibitem{Omura15}
Y.~{Omura}, Y.~{Miyashita}, M.~{Yoshikawa}, D.~{Summers}, M.~{Hikishima},
  Y.~{Ebihara}, Y.~{Kubota}, {Formation process of relativistic electron flux
  through interaction with chorus emissions in the Earth's inner
  magnetosphere}, \jgr 120 (2015) 9545--9562.
\newblock \href {http://dx.doi.org/10.1002/2015JA021563}
  {\path{doi:10.1002/2015JA021563}}.

\bibitem{Osmane16}
A.~{Osmane}, L.~B. {Wilson}, III, L.~{Blum}, T.~I. {Pulkkinen}, {On the
  Connection between Microbursts and Nonlinear Electronic Structures in
  Planetary Radiation Belts}, \apj 816 (2016) 51.
\newblock \href {http://dx.doi.org/10.3847/0004-637X/816/2/51}
  {\path{doi:10.3847/0004-637X/816/2/51}}.

\bibitem{Hsieh&Omura17}
Y.-K. {Hsieh}, Y.~{Omura}, {Nonlinear dynamics of electrons interacting with
  oblique whistler mode chorus in the magnetosphere}, \jgr 122 (2017) 675--694.
\newblock \href {http://dx.doi.org/10.1002/2016JA023255}
  {\path{doi:10.1002/2016JA023255}}.

\bibitem{Shklyar17}
D.~R. Shklyar, {Energy transfer
  from lower energy to higher-energy electrons mediated by whistler waves in
  the radiation belts}, \jgr 122~(1) (2017) 640--655.
\newblock \href {http://dx.doi.org/10.1002/2016JA023263}
  {\path{doi:10.1002/2016JA023263}}.

\bibitem{Demekhov17}
A.~G. {Demekhov}, U.~{Taubenschuss}, O.~{Santol{\'{\i}}k}, {Simulation of VLF
  chorus emissions in the magnetosphere and comparison with THEMIS spacecraft
  data}, \jgr 122 (2017) 166--184.
\newblock \href {http://dx.doi.org/10.1002/2016JA023057}
  {\path{doi:10.1002/2016JA023057}}.

\bibitem{Artemyev18:cnsns}
A.~V. {Artemyev}, A.~I. {Neishtadt}, D.~L. {Vainchtein}, A.~A. {Vasiliev},
  I.~Y. {Vasko}, L.~M. {Zelenyi}, {Trapping (capture) into resonance and
  scattering on resonance: Summary of results for space plasma systems},
  Communications in Nonlinear Science and Numerical Simulations 65 (2018)
  111--160.
\newblock \href {http://dx.doi.org/10.1016/j.cnsns.2018.05.004}
  {\path{doi:10.1016/j.cnsns.2018.05.004}}.

\bibitem{Shklyar09:review}
D.~{Shklyar}, H.~{Matsumoto}, {Oblique Whistler-Mode Waves in the Inhomogeneous
  Magnetospheric Plasma: Resonant Interactions with Energetic Charged
  Particles}, Surveys in Geophysics 30 (2009) 55--104.
\newblock \href {http://dx.doi.org/10.1007/s10712-009-9061-7}
  {\path{doi:10.1007/s10712-009-9061-7}}.

\bibitem{Artemyev14:pop}
A.~V. {Artemyev}, A.~A. {Vasiliev}, D.~{Mourenas}, O.~V. {Agapitov}, V.~V.
  {Krasnoselskikh}, {Electron scattering and nonlinear trapping by oblique
  whistler waves: The critical wave intensity for nonlinear effects}, Physics
  of Plasmas 21~(10) (2014) 102903.
\newblock \href {http://dx.doi.org/10.1063/1.4897945}
  {\path{doi:10.1063/1.4897945}}.

\bibitem{Albert13:AGU}
J.~M. {Albert}, X.~{Tao}, J.~{Bortnik}, {Aspects of Nonlinear Wave-Particle
  Interactions}, in: D.~{Summers}, I.~U. {Mann}, D.~N. {Baker}, M.~{Schulz}
  (Eds.), Dynamics of the Earth's Radiation Belts and Inner Magnetosphere,
  American Geophysical Union, 2013.
\newblock \href {http://dx.doi.org/10.1029/2012GM001324}
  {\path{doi:10.1029/2012GM001324}}.

\bibitem{Omura07}
Y.~{Omura}, N.~{Furuya}, D.~{Summers}, {Relativistic turning acceleration of
  resonant electrons by coherent whistler mode waves in a dipole magnetic
  field}, \jgr 112 (2007) 6236.
\newblock \href {http://dx.doi.org/10.1029/2006JA012243}
  {\path{doi:10.1029/2006JA012243}}.

\bibitem{Demekhov09}
A.~G. {Demekhov}, V.~Y. {Trakhtengerts}, M.~{Rycroft}, D.~{Nunn}, {Efficiency
  of electron acceleration in the Earth's magnetosphere by whistler mode
  waves}, Geomagnetism and Aeronomy 49 (2009) 24--29.
\newblock \href {http://dx.doi.org/10.1134/S0016793209010034}
  {\path{doi:10.1134/S0016793209010034}}.

\bibitem{Artemyev16:ssr}
A.~Artemyev, O.~Agapitov, D.~Mourenas, V.~Krasnoselskikh, V.~Shastun, F.~Mozer,
  {Oblique whistler-mode
  waves in the earth's inner magnetosphere: Energy distribution, origins, and
  role in radiation belt dynamics}, Space Science Reviews 200 (2016) 261--355.
\newblock \href {http://dx.doi.org/10.1007/s11214-016-0252-5}
  {\path{doi:10.1007/s11214-016-0252-5}}.

\bibitem{bookAKN06}
V.~I. {Arnold}, V.~V. {Kozlov}, A.~I. {Neishtadt}, Mathematical Aspects of
  Classical and Celestial Mechanics, 3rd Edition, Dynamical Systems III.
  Encyclopedia of Mathematical Sciences, Springer-Verlag, New York, 2006.

\bibitem{Artemyev18:jpp}
A.~V. {Artemyev}, A.~I. {Neishtadt}, A.~A. {Vasiliev}, D.~{Mourenas},
  {Long-term evolution of electron distribution function due to nonlinear
  resonant interaction with whistler mode waves}, Journal of Plasma Physics 84
  (2018) 905840206.
\newblock \href {http://dx.doi.org/10.1017/S0022377818000260}
  {\path{doi:10.1017/S0022377818000260}}.

\bibitem{Zhang18:jgr:intensewaves}
X.-J. {Zhang}, R.~{Thorne}, A.~{Artemyev}, D.~{Mourenas}, V.~{Angelopoulos},
  J.~{Bortnik}, C.~A. {Kletzing}, W.~S. {Kurth}, G.~B. {Hospodarsky},
  {Properties
  of intense field-aligned lower-band chorus waves: Implications for nonlinear
  wave-particle interactions}, \jgr 123~(7)  5379--5393.
\newblock \href
  {http://arxiv.org/abs/https://agupubs.onlinelibrary.wiley.com/doi/pdf/10.1029/2018JA025390}
  {\path{arXiv:https://agupubs.onlinelibrary.wiley.com/doi/pdf/10.1029/2018JA025390}},
  \href {http://dx.doi.org/10.1029/2018JA025390}
  {\path{doi:10.1029/2018JA025390}}.

\bibitem{Mourenas18:jgr}
D.~{Mourenas}, X.-J. {Zhang}, A.~V. {Artemyev}, V.~{Angelopoulos}, R.~M.
  {Thorne}, J.~{Bortnik}, A.~I. {Neishtadt}, A.~A. {Vasiliev}, {Electron
  Nonlinear Resonant Interaction With Short and Intense Parallel Chorus Wave
  Packets}, \jgr 123 (2018) 4979--4999.
\newblock \href {http://dx.doi.org/10.1029/2018JA025417}
  {\path{doi:10.1029/2018JA025417}}.

\bibitem{Artemyev16:pop:letter}
A.~V. {Artemyev}, A.~I. {Neishtadt}, A.~A. {Vasiliev}, D.~{Mourenas}, {Kinetic
  equation for nonlinear resonant wave-particle interaction}, Physics of
  Plasmas 23~(9) (2016) 090701.
\newblock \href {http://dx.doi.org/10.1063/1.4962526}
  {\path{doi:10.1063/1.4962526}}.

\bibitem{Artemyev17:pre}
A.~V. {Artemyev}, A.~I. {Neishtadt}, A.~A. {Vasiliev}, D.~{Mourenas},
  {Probabilistic approach to nonlinear wave-particle resonant interaction},
  \pre 95~(2) (2017) 023204.
\newblock \href {http://dx.doi.org/10.1103/PhysRevE.95.023204}
  {\path{doi:10.1103/PhysRevE.95.023204}}.

\bibitem{Artemyev17:arXiv}
A.~V. {Artemyev}, A.~I. {Neishtadt}, A.~A. {Vasiliev}, D.~{Mourenas}, {Kinetic
  equation for systems with resonant captures and scatterings}, ArXiv
  e-prints\href {http://arxiv.org/abs/1710.04489} {\path{arXiv:1710.04489}}.

\bibitem{CourantHilbert}
R.~{Courant}, D.~{Hilbert}, {Methods of mathematical physics}, 1953.

\bibitem{Karpman&Shkliar77}
V.~I. {Karpman}, D.~R. {Shkliar}, {Particle precipitation caused by a single
  whistler-mode wave injected into the magnetosphere}, \planss 25 (1977)
  395--403.
\newblock \href {http://dx.doi.org/10.1016/0032-0633(77)90055-1}
  {\path{doi:10.1016/0032-0633(77)90055-1}}.

\bibitem{Kennel&Engelmann66}
C.~F. {Kennel}, F.~{Engelmann}, {Velocity Space Diffusion from Weak Plasma
  Turbulence in a Magnetic Field}, Physics of Fluids 9 (1966) 2377--2388.
\newblock \href {http://dx.doi.org/10.1063/1.1761629}
  {\path{doi:10.1063/1.1761629}}.

\bibitem{Shklyar81}
D.~R. {Shklyar}, {Stochastic motion of relativistic particles in the field of a
  monochromatic wave}, Sov. Phys. JETP 53 (1981) 1197--1192.

\bibitem{Albert93}
J.~M. {Albert}, {Cyclotron resonance in an inhomogeneous magnetic field},
  Physics of Fluids B 5 (1993) 2744--2750.
\newblock \href {http://dx.doi.org/10.1063/1.860715}
  {\path{doi:10.1063/1.860715}}.

\bibitem{Min14}
K.~{Min}, K.~{Liu}, W.~{Li}, {Signatures of electron Landau resonant
  interactions with chorus waves from THEMIS observations}, \jgr 119 (2014)
  5551--5560.
\newblock \href {http://dx.doi.org/10.1002/2014JA019903}
  {\path{doi:10.1002/2014JA019903}}.

\bibitem{Li17:Langmuir}
J.~{Li}, J.~{Bortnik}, X.~{An}, W.~{Li}, R.~M. {Thorne}, M.~{Zhou}, W.~S.
  {Kurth}, G.~B. {Hospodarsky}, H.~O. {Funsten}, H.~E. {Spence}, {Chorus Wave
  Modulation of Langmuir Waves in the Radiation Belts}, \grl 44 (2017) 11.
\newblock \href {http://dx.doi.org/10.1002/2017GL075877}
  {\path{doi:10.1002/2017GL075877}}.

\bibitem{Kurita18}
S.~Kurita, Y.~Miyoshi, S.~Kasahara, S.~Yokota, Y.~Kasahara, S.~Matsuda,
  A.~Kumamoto, A.~Matsuoka, I.~Shinohara,
  {Deformation
  of electron pitch angle distributions caused by upper-band chorus observed by
  the arase satellite}, Geophysical Research Letters 0~(ja).
\newblock \href
  {http://arxiv.org/abs/https://agupubs.onlinelibrary.wiley.com/doi/pdf/10.1029/2018GL079104}
  {\path{arXiv:https://agupubs.onlinelibrary.wiley.com/doi/pdf/10.1029/2018GL079104}},
  \href {http://dx.doi.org/10.1029/2018GL079104}
  {\path{doi:10.1029/2018GL079104}}.

\bibitem{OgawaLeonciniVasilievGarbet_preprint}
S.~{Ogawa}, X.~{Leoncini}, A.~{Vasiliev}, X.~{Garbet}, {Tailoring steep density
  profile with unstable points}, Physics Letters A (2018) in press \href
  {http://arxiv.org/abs/arXiv:1611.00063v3} {\path{arXiv:arXiv:1611.00063v3}},
  \href {http://dx.doi.org/10.1016/j.physd.2011.11.015}
  {\path{doi:10.1016/j.physd.2011.11.015}}.

\bibitem{Neishtadt99}
A.~I. {Neishtadt}, {On Adiabatic Invariance in Two-Frequency Systems}, in
  Hamiltonian Systems with Three or More Degrees of Freedom, ed. Simo C.,
  NATO ASI Series C. Dordrecht: Kluwer Acad. Publ. 533 (1999) 193--213.
\newblock \href {http://dx.doi.org/10.1063/1.166236}
  {\path{doi:10.1063/1.166236}}.

\bibitem{Neishtadt06}
A.~I. {Neishtadt}, A.~A. {Vasiliev}, {Destruction of adiabatic invariance at
  resonances in slow fast Hamiltonian systems}, Nuclear Instruments and Methods
  in Physics Research A 561 (2006) 158--165.
\newblock \href {http://arxiv.org/abs/arXiv:nlin/0511050}
  {\path{arXiv:arXiv:nlin/0511050}}, \href
  {http://dx.doi.org/10.1016/j.nima.2006.01.008}
  {\path{doi:10.1016/j.nima.2006.01.008}}.

\end{thebibliography}

\end{document}